# Analyzing the activities of visitors of the Leiden Ranking website

Nees Jan van Eck and Ludo Waltman

Centre for Science and Technology Studies, Leiden University, The Netherlands
{ecknjpvan, waltmanlr}@cwts.leidenuniv.nl

To provide a better understanding of the way in which university rankings are used, we present a detailed analysis of the activities of visitors of a university ranking website. We use the website of the CWTS Leiden Ranking for this purpose. We for instance study the countries from which visitors originate, the specific pages on the Leiden Ranking website that they visit, the countries or the universities that they find of special interest, and the indicators that they focus on. In addition, we also discuss two experiments that were carried out on the Leiden Ranking website. Our analysis does not only provide new insights into the use of university rankings, but it also suggests possible ways in which these rankings can be improved.

## 1. Introduction

In the scientometric literature, university rankings are discussed primarily from a methodological point of view (e.g., Billaut, Bouyssou, & Vincke, 2010; Bookstein, Seidler, Fieder, & Winckler, 2010; Dehon, McCathie, & Verardi, 2010; Saisana, d'Hombres, & Saltelli, 2011; Van Raan, 2005; Waltman et al., 2012; Zitt & Filliatreau, 2007). In this paper, we take a different perspective. In our view, constructing a high-quality university ranking requires not only an advanced understanding of methodological issues but also a sufficient level of knowledge of the way in which university rankings are used. The use of university rankings has been studied using questionnaires and interviews (e.g., Hazelkorn, 2015). We take an alternative approach by analyzing the activities of visitors of a university ranking website. For this purpose, we use the website of the CWTS Leiden Ranking (LR), a university ranking produced by our center.

By analyzing the activities of visitors of the LR website, we intend to make two contributions. First, we aim to obtain a better understanding of the use of university rankings: Who is visiting university ranking websites, and what are visitors interested in? For instance, which countries or which universities do visitors find of special interest, and which indicators do they focus on? Our findings are specific for the LR, but we expect that to some extent they are also representative for university rankings more generally. Second, based on information about the use of university rankings, we aim to learn more about possible ways in which these rankings can be improved. Improvements may for instance relate to the information that is made available in a ranking and the way in which this information is presented.

The LR is available at [www.leidenranking.com](www.leidenranking.com). The ranking provides bibliometric indicators for almost 1000 major universities worldwide. Starting from 2012, each year a new edition of the LR has been released by our center, the Centre for Science and Technology Studies (CWTS) at Leiden University. The 2018 edition currently is the most recent one. We refer to Waltman et al. (2012) for an introduction to the LR. Although the description of the LR provided by Waltman et al. (2012) is not entirely up-to-date anymore, the paper still offers a useful overview of the general philosophy of the ranking.

In the first editions of the LR, the focus was on improving the ranking by increasing the number of universities that are included, by refining the data collection methodology, and by extending and improving the bibliometric indicators that are made available. In recent years, the focus has changed and a significant amount of effort has been put into improving the online presentation of the LR and providing guidelines for proper use of university rankings in general and the LR in particular (Waltman, Wouters, & Van Eck, 2017). We are now shifting our attention to analyzing how the LR is used.

Our analysis focuses on the 2017 edition of the LR. We study how visitors make use of the website of the LR 2017. The LR 2017 was released on May 17, 2017. Between May 17, 2017 and February 28, 2018, the activities of visitors of the LR 2017 website were recorded. Our analysis is based on the activities that took place during this period. In addition, we also discuss two experiments that were carried out on the LR 2017 website.

The rest of this paper is organized as follows. In Section 2, we discuss the collection of the data on which our analysis is based. In Section 3, we present the results of the analysis. We summarize our conclusions in Section 4.

## 2. Data

The 2017 edition of the LR was released on May 17, 2017 at 13h CEST. Starting from the release of the LR 2017, the activities of visitors of the LR website were recorded. More precisely, the activities on the following three web pages were recorded:

- List view page: www.leidenranking.com/ranking/2017/list
- Chart view page: www.leidenranking.com/ranking/2017/chart
- Map view page: www.leidenranking.com/ranking/2017/map

These pages provide three different perspectives on the LR, referred to as the list view, the chart view, and the map view, respectively (see Figure 1). The list view presents universities in a list ordered based on a bibliometric indicator. The chart view presents universities in a scatter plot, with one bibliometric indicator on the horizontal axis and another bibliometric indicator on the vertical axis. The map view takes a geographical perspective. It shows universities in a world map. In addition to the three pages mentioned above, there is also a university page (see Figure 1). This page provides detailed statistics at the level of an individual university. Activities on this page were recorded as well. The analysis presented in this paper is based on activities that were recorded between May 17, 2017 and February 28, 2018.

When someone visited the four web pages discussed above, this was recorded. In addition, each time a visitor performed an action, this was recorded as well. Performing an action means that a visitor moves from one page to another or changes a setting on a page (e.g., changing the currently selected time period, field, country, or indicator). When multiple actions are performed consecutively in the same browser window, these actions are part of the same browser session. The actions are also referred to as views. Each session consists of one or more views.

For each visitor, an IP address is available. Based on the IP address, the country of a visitor was determined. We used the MaxMind GeoLite database (http://dev.maxmind.com/geoip/geolite) for this purpose.

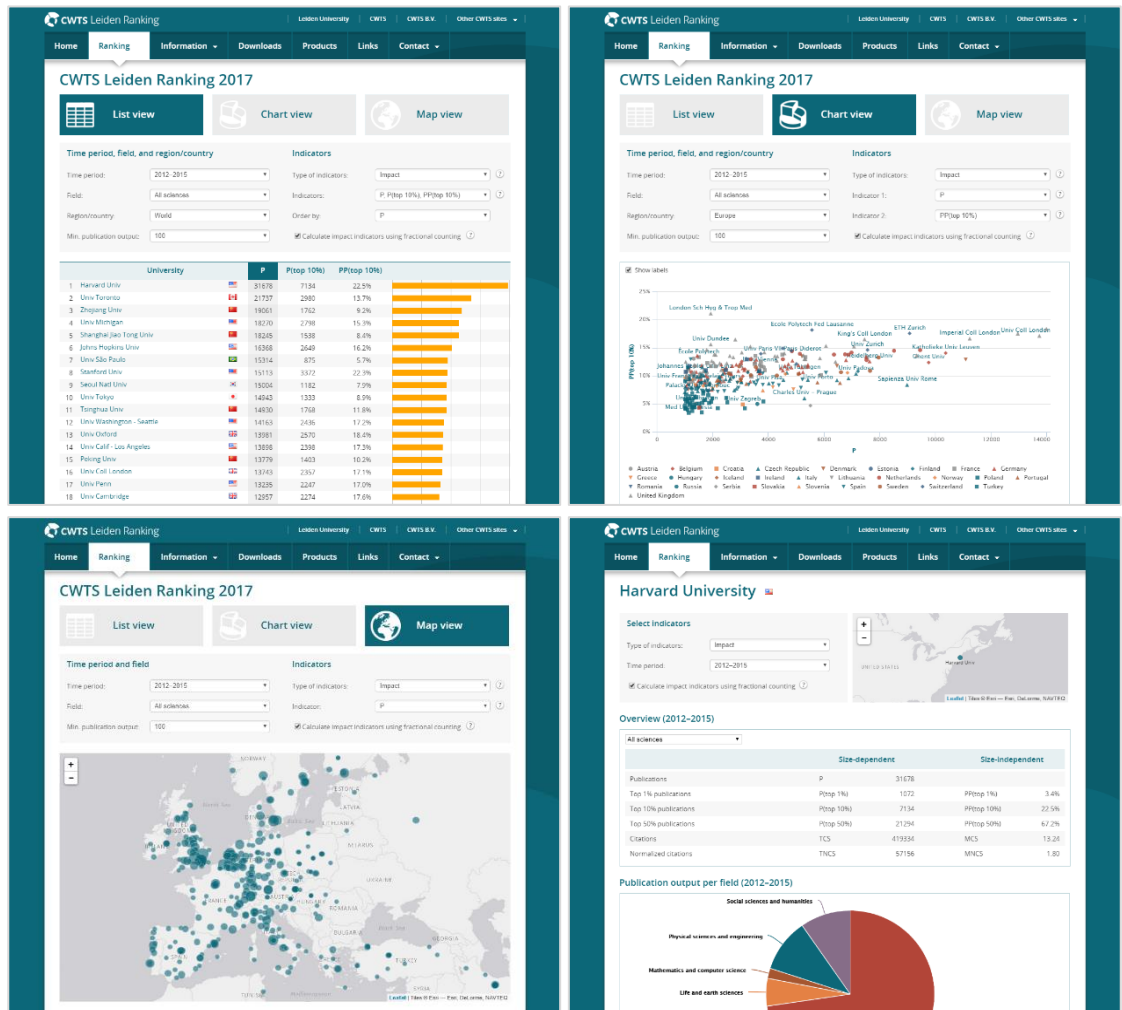

Figure 1. The list view page (top left), the chart view page (top right), the map view page (bottom left), and the university page (bottom right).

Finally, we note that visits from Googlebot, the indexing spider of Google, were filtered out. No other non-human visitors were found that needed to be filtered out. However, we did filter out visits from IP addresses of CWTS.

To facilitate reproducibility and follow-up research, the data on which our analysis is based has been made publicly available (Van Eck & Waltman, 2018).

## 3. Results

We now present the results of our analysis. We first report results for the LR 2017 website as a whole in Subsection 3.1. We then present results for the list view page and the university page in Subsections 3.2 and 3.3, respectively. Finally, in Subsection 3.4, we discuss the results of two experiments that we performed.

**3.1. Leiden Ranking 2017 website**

In total, data was collected for 92,029 sessions. Hence, between May 17, 2017 and February 28, 2018, the LR 2017 website was visited 92,029 times, which corresponds with an average of 319.5 visits per day. Figure 2 shows for each month in the period of analysis the average daily number of visits. As may be expected, the LR website was visited most often in the month of the release of the 2017 edition. In May 2017, on average the website was visited almost 2,000 times per day (taking into account only the second half of the month, starting from the release of the LR 2017 on May 17). In later months, the average daily number of visits decreased, reaching a stable level of about 200 visits per day.

As discussed in Section 2, each session consists of one or more views. In our period of analysis, a session on average consisted of 4.7 views. Figure 3 shows the distribution of the number of views per session. As can be seen, the distribution is quite skewed. Of all sessions, 38.6% consisted of just one view, while 9.9% consisted of more than 10 views.

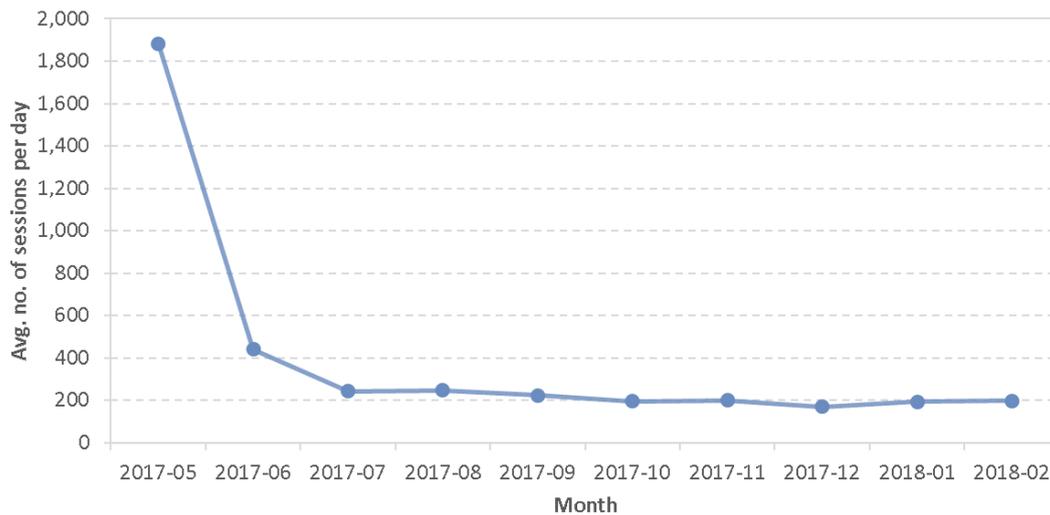

Figure 2. Time trend of the average number of sessions per day.

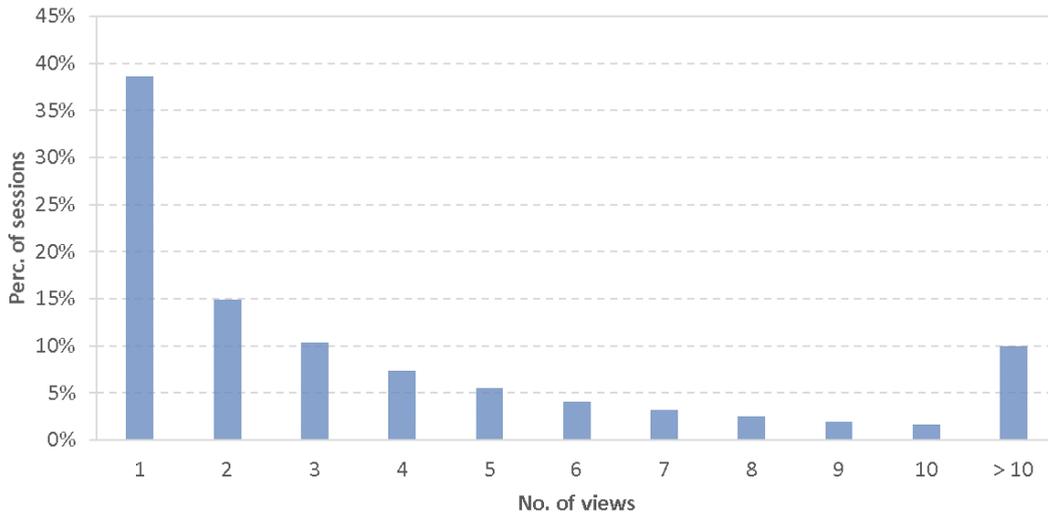

Figure 3. Distribution of the number of views per session.

For each session, we know the country from which the LR 2017 website is visited. In total, the LR 2017 website was visited from 185 countries. Table 1 lists the top 20 countries responsible for the largest number of sessions. For each country, the table reports the share of all sessions originating from this country. In total, the top 20 countries account for 79.0% of all sessions. Table 1 also shows for each country the average number of views per session and the number of universities included in the LR 2017. Not surprisingly, a large number of sessions (i.e., 6.0% of the total) originated from the Netherlands. In addition to Western European countries, it turns out that the US, Australia, Turkey, Iran, and South Korea account for a large number of sessions. The number of sessions originating from China is relatively limited, given the size of the Chinese research system and the number of Chinese universities included in the LR 2017. We further note that there are substantial differences between countries in the average number of views per session (e.g., 6.3 views per session for Sweden vs. 2.9 views per session for Taiwan), suggesting that visitors from some countries tend to study the LR in more detail than visitors from other countries.

In each session, one or more pages of the LR 2017 website were visited. As discussed in Section 2, there are four pages: the list view page, the chart view page, the map view page, and the university page. For each of these four pages, Table 2 reports the share of all sessions in which the page was visited at least once. In addition, for each page, the table also shows the average number of views of the page per session, including only sessions in which the page has at least one view.

Table 1. Top 20 countries responsible for the largest number of sessions.

| Country | Perc. of sessions | Avg. no. of views per session | No. of universities in the LR 2017 |
|---|---|---|---|
| United States | 9.6% | 4.1 | 177 |
| Australia | 6.0% | 4.7 | 25 |
| Netherlands | 6.0% | 4.6 | 13 |
| United Kingdom | 5.8% | 5.0 | 47 |
| Turkey | 5.6% | 3.9 | 16 |
| Iran | 5.5% | 5.8 | 18 |
| South Korea | 5.4% | 5.8 | 35 |
| France | 5.2% | 4.4 | 24 |
| Germany | 3.7% | 6.0 | 50 |
| Denmark | 3.6% | 4.6 | 5 |
| Switzerland | 3.4% | 3.8 | 7 |
| Spain | 3.3% | 5.4 | 34 |
| Canada | 3.0% | 5.0 | 28 |
| China | 2.8% | 3.9 | 138 |
| Portugal | 2.4% | 5.9 | 6 |
| Japan | 1.8% | 3.8 | 41 |
| Italy | 1.7% | 5.8 | 39 |
| Taiwan | 1.5% | 2.9 | 17 |
| India | 1.3% | 4.7 | 20 |
| Sweden | 1.3% | 6.3 | 10 |

As can be seen in Table 2, visitors of the LR 2017 website spent most of their time on the list view page. This page was visited in 92.5% of all sessions, and the average number of views was substantially higher than for the other pages. Hence, the statistics presented in Table 2 seem to indicate that visitors of the LR 2017 website are interested mainly in the list view. However, to some extent this may also be an artifact, since the list view is the default view presented to visitors of the LR 2017 website. In any case, it is clear that the chart view page, the map view page, and the university page were visited much less often than the list view page. For this reason, our focus is mostly on the list view page in the remainder of this section.

Table 2. Share of all sessions in which the different pages of the LR 2017 website were visited.

| Page | Perc. of sessions | Avg. no. of views per session |
|---|---|---|
| List view | 92.5% | 4.1 |
| Chart view | 10.3% | 2.2 |
| Map view | 10.5% | 1.8 |
| University | 23.1% | 2.3 |

**3.2. List view page**

We now focus specifically on the list view page. We consider only sessions in which this page was visited.

Table 3 lists the settings that can be changed by a visitor of the list view page. For each of these settings, Figure 4 shows the share of all sessions in which the setting was changed. The field and the region/country settings were changed in about one-third of all sessions. The order by setting, which determines the indicator based on which universities are ordered, was changed in 17.4% of all sessions. Hence, in somewhat more than one-sixth of all sessions, visitors choose to switch from the default ordering of universities based on publication output to an alternative ordering based on a different indicator. As can be seen in Figure 4, the other settings available on the list view page were changed less frequently. The setting that was changed least often is the counting method setting. In only 4.0% of all sessions, visitors choose to switch from the default fractional counting method to the full counting method (for more information about the difference between the two counting methods, see Waltman & Van Eck, 2015).

For each of the five broad fields of science distinguished in the LR 2017, Figure 5 shows the share of all sessions in which the field was selected. The differences are not very large, with the most popular field, physical sciences and engineering, being selected less than twice as often as the least popular field, life and earth sciences.

Table 3. Overview of the settings that can be changed on the list view page.

| Setting | Description | Default choice |
|---|---|---|
| Time period | Choice of a time period | 2012–2015 |
| Field | Choice of a field of science | All sciences |
| Region/country | Choice of a region (i.e., continent) or a country | World |
| Min. publication output | Choice of the minimum publication output that a university is required to have | 100 |
| Type of indicators | Choice between impact (citation) and collaboration (co-authorship) indicators | Impact |
| Indicators | Choice of specific impact or collaboration indicators | P, P(top 10%), PP(top 10%) |
| Order by | Choice of the indicator based on which universities are ordered; universities can also be ordered alphabetically based on their name | P |
| Counting method | Choice between full and fractional counting | Fractional counting |

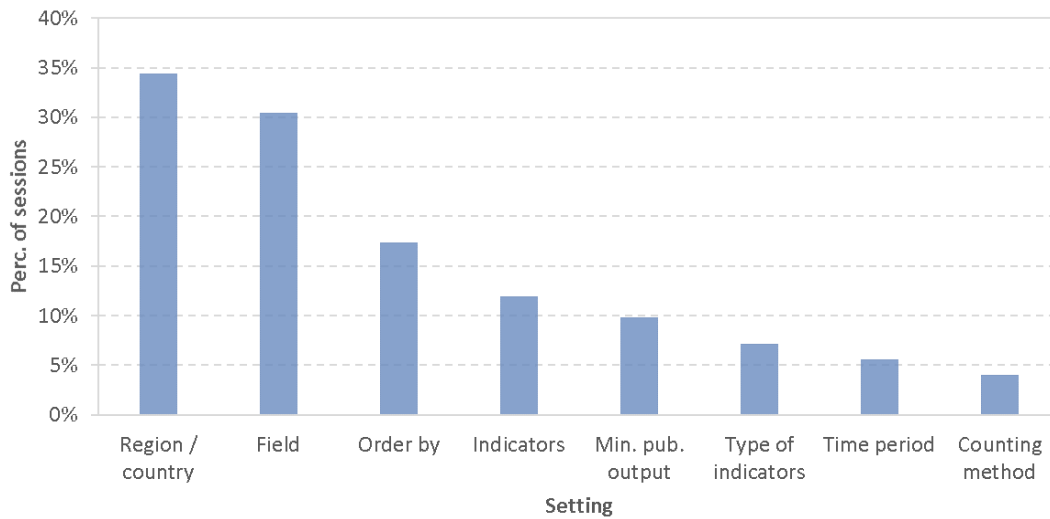

Figure 4. Share of all sessions in which a specific setting was changed on the list view page.

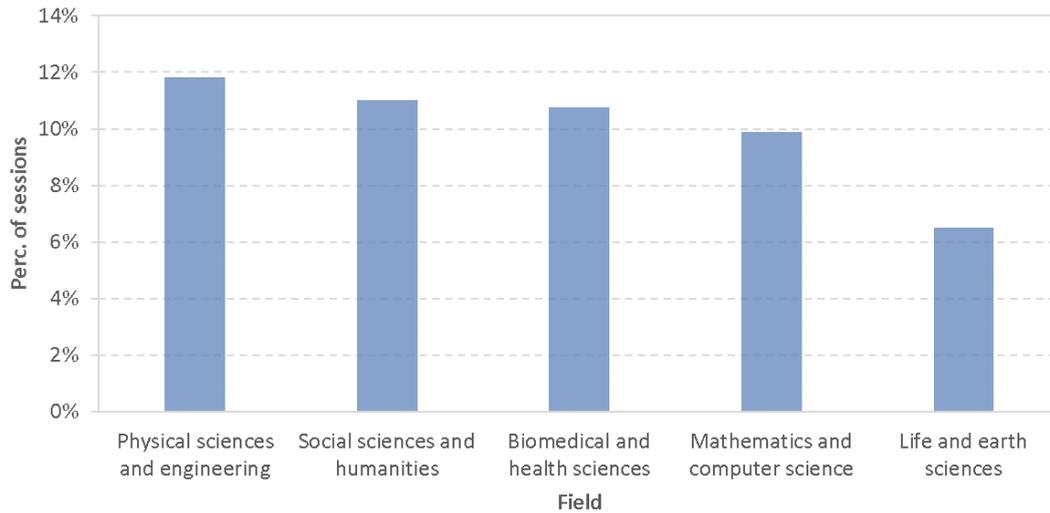

Figure 5. Share of all sessions in which a specific field was selected on the list view page.

Figure 6 shows the share of all sessions in which a specific region (i.e., continent) was selected. Similar statistics are reported in Figure 7 at the level of countries instead of regions. Europe is by far the most popular region. It was selected in 10.1% of all sessions, while each of the other regions was selected in less than 3% of the sessions. Nevertheless, of the five most popular countries, three (i.e., Iran, South Korea, and Australia) are located outside Europe.

Since we know the country of each visitor, we were able to determine how frequently visitors from a specific country are interested in universities either in their own country or in other countries. We counted for each visiting country the number of sessions in which visitors from that country selected a specific country on the list view page. For the top 10 visiting countries and the top 10 countries that were selected most often on the list view page, Figure 8 presents an alluvial diagram that shows the relations between visiting countries and countries selected on the list view page. Not surprisingly, visitors have a strong interest in universities in their own country. However, a few significant relations between different countries are visible as well. In particular, visitors from Turkey have a strong interest in UK universities. Also, visitors from Iran are relatively strongly interested in German universities.

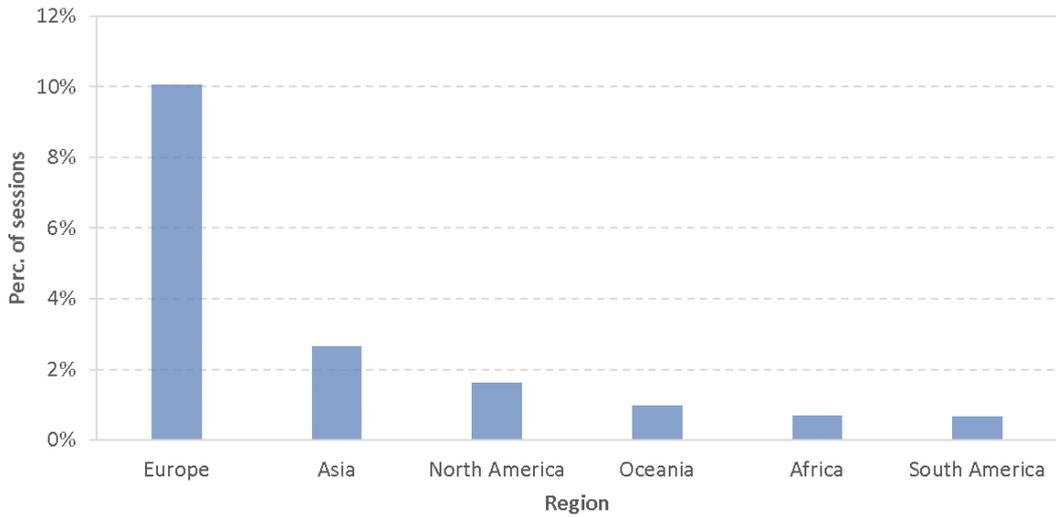

Figure 6. Share of all sessions in which a specific region was selected on the list view page.

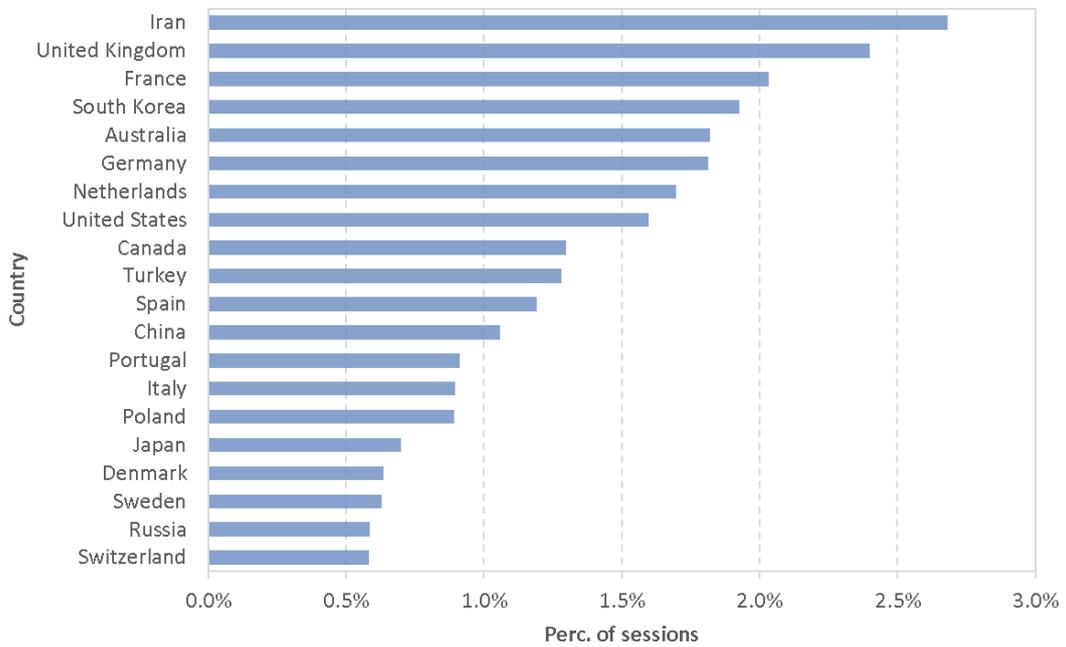

Figure 7. Share of all sessions in which a specific country was selected on the list view page (top 20 countries only).

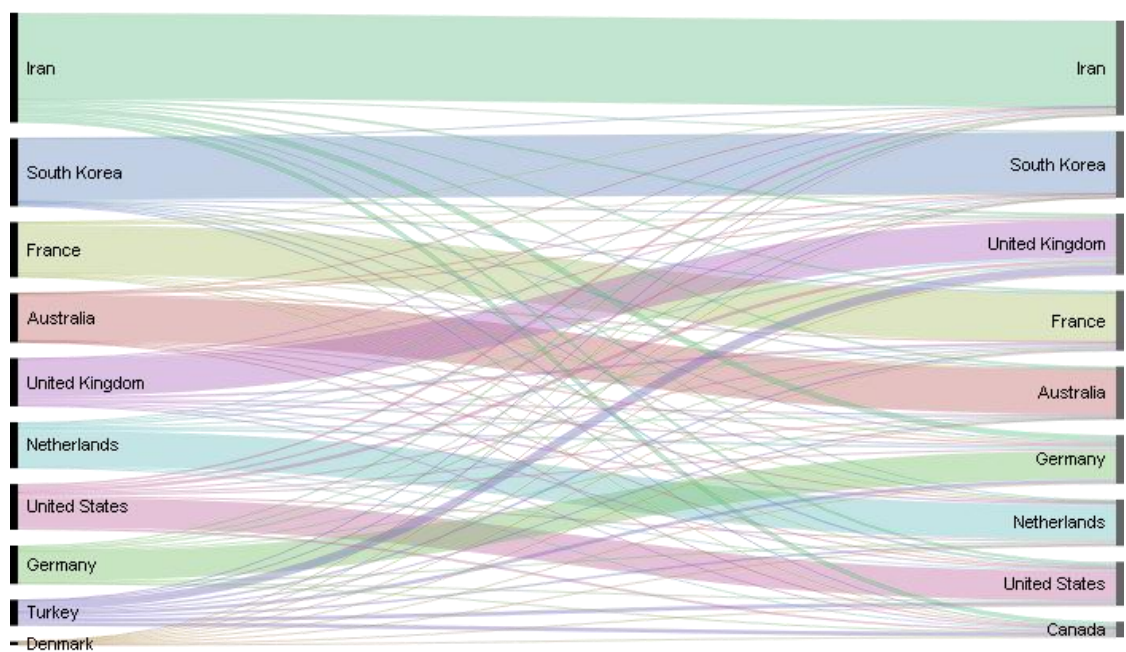

Figure 8. Alluvial diagram of the relations (in terms of numbers of sessions) between the top 10 visiting countries (on the left) and the top 10 countries selected most often on the list view page (on the right).

By default, the list view presents indicators of scientific impact. As can be seen in Figure 4, in only 7.2% of all sessions, the type of indicators setting was changed. Hence, visitors choose to switch from indicators of scientific impact (based on citations) to indicators of scientific collaboration (based on co-authorships) only in a small share of all sessions. This is also visible in Figure 9, which shows the share of all sessions in which a specific indicator for ordering universities was selected (for more information about the indicators that are available in the LR, see www.leidenranking.com/information/indicators/). Each of the collaboration indicators was selected only in a very small share of all sessions. The PP(int collab) indicator (i.e., the proportion of internationally collaborative publications) is the collaboration indicator that was selected most often, but even this indicator was selected in only 0.8% of all sessions.

As we have seen in Figure 4, in about one-sixth of all sessions, visitors choose to switch from the default ordering of universities based on publication output (i.e., the P indicator) to an alternative ordering based on a different indicator. Figure 9 shows that visitors are more interested in size-independent indicators, labeled as PP(...) indicators, than in size-dependent indicators, labeled as P(...) indicators. Size-independent indicators (e.g., the proportion of highly cited publications of a university) provide a relative perspective on the performance of a university, that is, a perspective that has been corrected for university size, where university size is quantified by the total publication output of a university. On the other hand, size-dependent indicators (e.g., the total number of highly cited publications of a university) offer an absolute perspective on the performance of a university, that is, a perspective in which no correction has been made for university size. As can be seen in Figure 9, for each size-independent indicator, the share of all sessions in which the indicator was selected is higher than the share of all sessions in which the corresponding size-dependent indicator was selected. We note that Figure 9 also shows that ordering universities alphabetically based on their name is a relatively popular option.

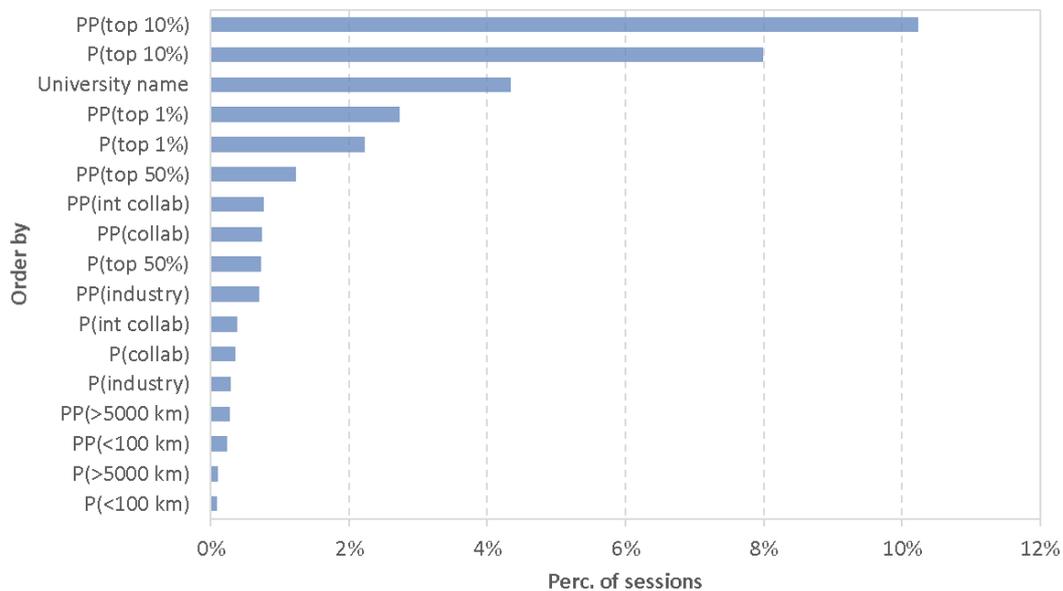

Figure 9. Share of all sessions in which a specific indicator for ordering universities was selected.

**3.3. University page**

We now briefly consider the university page. As can be seen in Table 2, the university page was visited in less than one quarter of all sessions. We therefore do not present detailed results for this page. However, we do discuss the universities that received most attention.

Table 4. Top 20 universities for which the university page was visited most often.

| University | Country | Perc. of sessions |
|---|---|---|
| Harvard University | United States | 3.3% |
| University of Toronto | Canada | 1.9% |
| Zhejiang University | China | 1.8% |
| University of Paris VI - Pierre and Marie Curie | France | 1.6% |
| University of São Paulo | Brazil | 1.6% |
| Utrecht University | Netherlands | 1.5% |
| University of Copenhagen | Denmark | 1.4% |
| University of Lisbon | Portugal | 1.3% |
| Islamic Azad University Science & Research Tehran | Iran | 1.1% |
| Ulsan National Institute of Science and Technology | South Korea | 1.1% |
| University of British Columbia | Canada | 1.1% |
| Rockefeller University | United States | 1.1% |
| Tsinghua University | China | 1.1% |
| Babeș-Bolyai University | Romania | 1.0% |
| ETH Zurich | Switzerland | 1.0% |
| University of Queensland | Australia | 1.0% |
| University College London | United Kingdom | 1.0% |
| Leiden University | Netherlands | 1.0% |
| University of Cape Town | South Africa | 0.9% |
| Seoul National University | South Korea | 0.9% |

Table 4 lists the top 20 universities for which the university page was visited most often. The three universities visited in the largest number of sessions are Harvard University, University of Toronto, and Zhejiang University. These are also the three universities with the largest publication output in the LR 2017. In the list view, universities are by default ordered based on the P indicator (i.e., publication output), which means that Harvard University, University of Toronto, and Zhejiang University are listed first, second, and third, respectively. Most likely, this is an important reason why the university page was visited so often for these three universities. Rockefeller University is also included in Table 4. This is the university that is listed first in the list view when universities are ordered based on the PP(top 10%) indicator (i.e., the proportion of publications belonging to the top 10% most cited of their field and publication year). Various other universities seem to be included in Table 4 because they are listed first in the list view when only universities from a specific country are considered. Of course, some universities may be included in Table 4 for very different reasons, for instance because they pay a lot of attention to the LR and perhaps also actively promote the ranking.

**3.4. Experiments**

An important element in the philosophy of the LR is that from the point of view of CWTS there is no best indicator for ranking universities. The choice of the indicator that is used to rank universities should be dependent on the purpose for which the LR is used. Different users may use the LR for different purposes and may therefore prefer to use different indicators for ranking universities. However, when the list view page is visited, there needs to be a default criterion for ordering universities. In earlier editions of the LR, the PP(top 10%) indicator was used as the default criterion. Based on the idea that we do not want to suggest a preference for either size-independent or size-dependent indicators of scientific impact, the default criterion was changed to the P indicator in the 2016 edition of the LR.

In practice, we have the impression that the default criterion for ordering universities on the list view page is often perceived as the criterion that is recommended by CWTS as the best criterion for ranking universities. Various reports in the popular press seem to be based on this incorrect perception. Also, the fact that the order by setting on the list view page was changed in only 17.4% of all sessions (see Figure 4) shows that many visitors of the list view page stick to the default criterion for ordering universities, probably either because they consider this to be the criterion that is recommended by CWTS or because they are not even aware that it is possible to select an alternative criterion. Hence, even though CWTS does not want to recommend a specific criterion for ranking universities, the default criterion used on the list view page seems to be perceived by many users of the LR as the criterion that is recommended by CWTS.

We performed two experiments in order to try to make visitors of the LR website more aware that they need to decide themselves how they want universities to be ranked. In experiment 1, the P indicator was maintained as the default criterion for ordering universities on the list view page, but the following message was prominently displayed when the default criterion was used: "By default universities are ordered based on the P indicator (number of publications). Make sure to select your preferred indicator for ordering the universities." In experiment 2, by default universities were ordered alphabetically based on their name. In addition, the following message was displayed: "By default universities are ordered alphabetically based on their name. Make sure to select your preferred indicator for ordering the universities."

Experiment 1 took place between March 1 and March 8, 2018. In this period, there were 1,395 sessions in which the list view page was visited, with an average of 4.2 views per session (just slightly above the average of 4.1 views per session reported in Table 2). In 22.4% of all sessions, visitors choose to switch from the default criterion for ordering universities (i.e., the P indicator) to an alternative criterion. For comparison, in Subsection 3.2, we found that the order by setting was changed in 17.4% of all sessions. Figure 10 shows the top 5 indicators that were selected most often for ordering universities. This top 5 is identical to the top 5 in Figure 9. However, for each of the indicators, the share of all sessions in which the indicator was selected is somewhat higher. For instance, comparing Figure 10 to Figure 9, the share of all sessions in which the PP(top 10%) indicator was selected shows an increase from 10.2% to 11.9%.

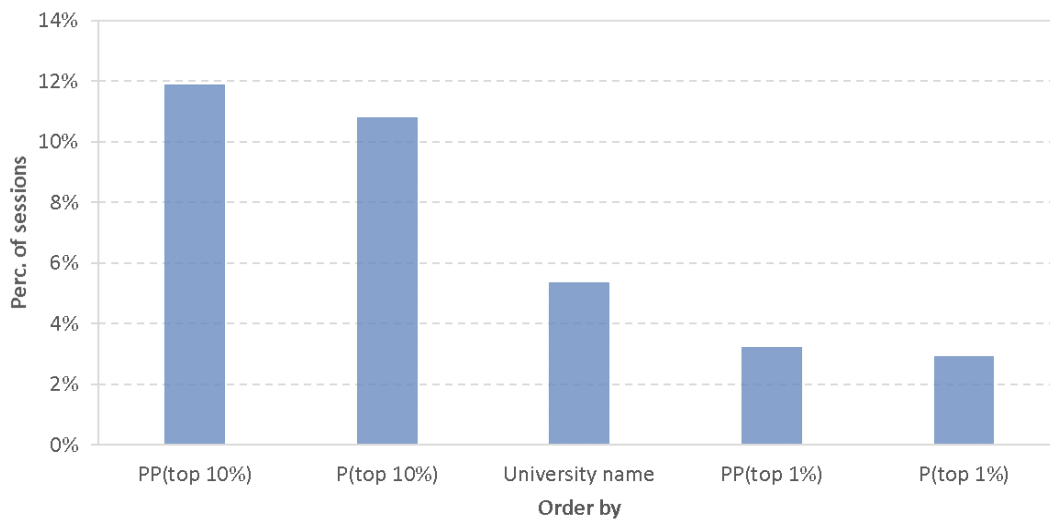

Figure 10. Share of all sessions in experiment 1 in which a specific indicator for ordering universities was selected (top 5 indicators only).

Experiment 2 was carried out between March 12 and March 19, 2018. The list view page was visited in 1,571 sessions in this period. The average number of views per session was 4.7, which is substantially higher than the average of 4.2 views per session in experiment 1. Moreover, in 45.0% of all sessions, visitors choose to switch from the default criterion for ordering universities (i.e., university name) to an alternative criterion. As can be seen in Figure 11, the indicator that was selected most often for ordering universities is the P indicator, followed by the P(top 10%) and PP(top 10%) indicators.

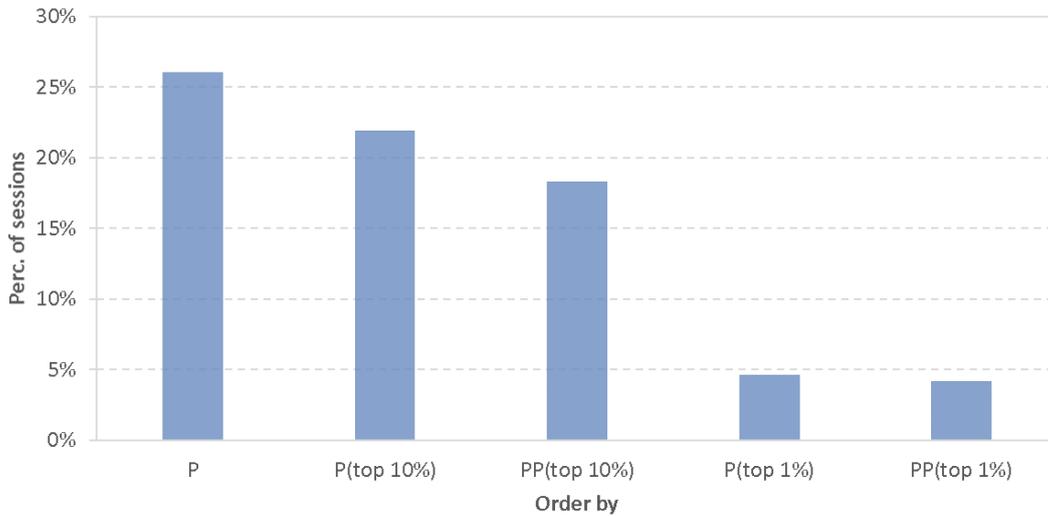

Figure 11. Share of all sessions in experiment 2 in which a specific indicator for ordering universities was selected (top 5 indicators only).

## 4. Conclusions

To guide the construction of university rankings, it is important to understand how these rankings are used. The use of university rankings can be studied in various ways. In this paper, we have analyzed the activities of visitors of the LR website. To the best of our knowledge, this is the first paper analyzing the activities of visitors of a university ranking website.

Based on our analysis, the observations that we consider most interesting can be summarized as follows:

1. Some countries account for a disproportionally large share of all visitors of the LR website. Many visitors originate from European countries. Outside Europe, the large number of visitors from Australia, Iran, and South Korea is remarkable.[1] On the other hand, the number of visitors from certain other countries, such as China, is relatively small. It is not immediately clear why visitors from certain countries are overrepresented. These countries may have a specific interest in the LR, but presumably they have a strong interest in university rankings in general.

---

[1] Some background information on the use of university rankings in Australia is provided by Hazelkorn (2009). For South Korea, we refer to Yonezawa, Chen, Jung, and Lo (2017) for background information on the use of university rankings.

2. Visitors of the LR website pay much more attention to the list view than to the chart view and the map view. Probably this is partly because the list view is presented as the default view on the LR website. However, based on our contacts with users of the LR, we also have the impression that many users of the ranking are attracted by the simplicity of the list view. In addition, of the three views provided on the LR website, the list view of course matches best with the traditional idea of a university ranking as a ranked list of universities.

3. Visitors of the LR website do not pay much attention to indicators of scientific collaboration. Indicators of scientific impact are much more popular.

4. Visitors of the LR website are more interested in size-independent indicators than in size-dependent indicators. However, the difference is not very large. This offers support for the way in which indicators are currently presented in the list view of the LR, with size-dependent and size-independent indicators consistently being reported together and without emphasizing one type of indicator over the other.

5. By default, universities are ordered based on publication output in the list view on the LR website. Visitors of the website usually do not change the criterion based on which universities are ordered. We are concerned that many visitors of the LR website may not realize that they should decide themselves which criterion they consider most appropriate for ordering universities. One of the experiments that we carried out seems to indicate that visitors of the website can be made more aware of this by changing the default ordering of universities. When universities are by default ordered alphabetically based on their name, visitors of the website are more likely to change the criterion for ordering universities.

We hope that the analysis presented in this paper will be useful in at least two ways. On the one hand, we hope to contribute to a better understanding of the use of university rankings. On the other hand, we hope that our analysis will help to improve university rankings. We see our work as part of a broader endeavor to systematically study the use of scientometric tools, relying on approaches ranging from usability testing to questionnaires and interviews.

## References

Billaut, J.-C., Bouyssou, D., & Vincke, P. (2010). Should you believe in the Shanghai ranking? An MCDM view. *Scientometrics*, *84*(1), 237–263.

Bookstein, F.L., Seidler, H., Fieder, M., & Winckler, G. (2010). Too much noise in the Times Higher Education rankings. *Scientometrics*, *85*(1), 295–299.

Dehon, C., McCathie, A., & Verardi, V. (2010). Uncovering excellence in academic rankings: A closer look at the Shanghai ranking. *Scientometrics*, *83*(2), 515–524.

Hazelkorn, E. (2009). Rankings and the battle for world-class excellence. *Higher Education Management and Policy*, *21*(1), 1–22.

Hazelkorn, E. (2015). *Rankings and the reshaping of higher education: The battle for world-class excellence*. Springer.

Saisana, M., d'Hombres, B., & Saltelli, A. (2011). Rickety numbers: Volatility of university rankings and policy implications. *Research Policy*, *40*(1), 165–177.

Van Raan, A.F.J. (2005). Fatal attraction: Conceptual and methodological problems in the ranking of universities by bibliometric methods. *Scientometrics*, *62*(1), 133–143.

Waltman, L., Calero-Medina, C., Kosten, J., Noyons, E.C.M., Tijssen, R.J.W., Van Eck, N.J., ... Wouters, P. (2012). The Leiden Ranking 2011/2012: Data collection, indicators, and interpretation. *Journal of the American Society for Information Science and Technology*, *63*(12), 2419–2432.

Waltman, L., & Van Eck, N.J. (2015). Field-normalized citation impact indicators and the choice of an appropriate counting method. *Journal of Informetrics*, *9*(4), 872–894.

Van Eck, N.J., & Waltman, L. (2018). *CWTS Leiden Ranking 2017 website* [Data set]. doi:10.17632/5bxw69mzht.1

Waltman, L., Wouters, P., & Van Eck, N.J. (2017, May 17). *Ten principles for the responsible use of university rankings* [Blog post]. Retrieved from www.cwts.nl/blog?article=n-r2q274

Yonezawa, A., Chen, S., Jung, J., & Lo, W.Y.W. (2017). East Asia: Catch-up and identity – developments in and impacts of university rankings. In E. Hazelkorn (Ed.), *Global rankings and the geopolitics of higher education: Understanding the* *influence and impact of rankings on higher education, policy and society* (pp. 116–127). Routledge.

Zitt, M., & Filliatreau, G. (2007). Big is (made) beautiful: Some comments about the Shanghai ranking of world-class universities. In J. Sadlak, & N.C. Liu (Eds.), *The world-class university and ranking: Aiming beyond status* (pp. 141–160). UNESCO-CEPES, Institute of Higher Education, Shanghai Jiao Tong University, and Cluj University Press.